\documentclass[12pt]{article}

\usepackage{amsmath}
\usepackage{amssymb}
\usepackage{amsfonts}
\usepackage{lscape}

\begin{document}

\fontsize{12}{6mm}\selectfont
\setlength{\baselineskip}{2em}

$~$\\[.35in]
\newcommand{\dss}{\displaystyle}
\newcommand{\raro}{\rightarrow}
\newcommand{\be}{\begin{equation}}

\def\sech{\mbox{\rm sech}}
\thispagestyle{empty}

\begin{center}
{\Large\bf Dynamics of Induced Surfaces in } \\    [2mm]
{\Large\bf Four-Dimensional Euclidean Space} \\    [2mm]
\end{center}

\vspace{1cm}
\begin{center}
{\bf Paul Bracken}                        \\
{\bf Department of Mathematics,} \\
{\bf University of Texas,} \\
{\bf Edinburg, TX  }  \\
{78541-2999}
\end{center}

\vspace{3cm}
\begin{abstract}
The Davey Stewartson hierarchy will be developed
based on a set of three matrix differential
operators. These equations will act as evolution
equations for different types of surface
deformation in Euclidean four space.
The Weierstrass representation for
surfaces will be developed and its 
uniqueness up to gauge transformations will
be reviewed. Applications of the hierarchy will be given 
with regard to generating deformations of surfaces,
and it will be shown that the Willmore
functional is preserved under this kind of
deformation.
\end{abstract}

\newpage
\begin{center}
{\bf 1. INTRODUCTION.}
\end{center}

\par
Surfaces and the dynamics of surfaces play a very
essential role in many areas of classical as well as
quantum physics. Moreover, as far as the area
of classical differential geometry is concerned, the
theory of the immersion and deformations of surfaces
has been the subject of intense research {\bf [1]}.

Domains of study, such as surface waves, deformation
of membranes, dynamics of vortex sheets as well as
certain problems in the area of hydrodynamics
are related to the motion of boundaries which
separate different regions. In particular,
in the area of string theory, the action is
related to the Polyakov integral over
surfaces. Certain special classes of surfaces
give important contributions to 
various types of physical
quantities appearing in these theories
and are of interest to consider.

Recently, Konopelchenko generalized the
Weierstrass formulas to the case of generic
surfaces in $\mathbb R^3$ {\bf [2-3]}. These formulas can be
used to study the global properties of surfaces in 
$\mathbb R^3$, as well as the integrable deformations
of such surfaces. This latter aspect is perhaps
one of the more important reasons for developing
these kinds of techniques for inducing surfaces 
in higher dimensional spaces which include Minkowski
type spaces as well as Euclidean spaces.
These representations can then be used to study
not only the geometry of surfaces, but the
integrable deformations of such surfaces as well.

The generalization of the Weierstrass formulas
to generic surfaces in $\mathbb R^3$ which was
proposed by Konopelchenko consists of the linear
system of Dirac equations
$$
\partial \psi = p \varphi,
\qquad
\bar{\partial} \varphi = -p \psi,
\eqno(1.1)
$$
where $\psi$ and $\varphi$ are complex-valued
functions of $z$, $\bar{z} \in \mathbb C$ and
$p ( z , \bar{z})$ is a real-valued function.
The derivative operators will be abbreviated to
$\partial = \partial / \partial z$ and 
$\bar{\partial} = \partial / \partial \bar{z}$
throughout.
Using solutions to (1.1), three real-valued functions
$X^1 ( z, \bar{z})$, $X^2 (z, \bar{z})$ and
$X^3 (z, \bar{z})$ are defined by the integrals
$$
X^1 + i X^2 =i \int_{\Gamma} ( \bar{\psi}^2 \, dz'
- \bar{\varphi}^2 \, d \bar{z}' ),
$$
$$
X^1 -i X^2 = i \int_{\Gamma} ( \varphi^2 \, dz' -
\psi^2 \, d \bar{z}'),
\eqno(1.2)
$$
$$
X^3 = - \int_{\Gamma} ( \bar{\psi} \varphi \, dz'
+ \psi \bar{\varphi} \, d \bar{z}'),
$$
where $\Gamma$ is an arbitrary curve in $\mathbb C$.
Regarding the $X^i (z, \bar{z})$ as coordinates
in $\mathbb R^3$, (1.1) and (1.2) define a conformal
immersion of a surface into $\mathbb R^3$.
It has been shown that the generalized
Weierstrass representation of Konopelchenko has several
other links to areas of mathematical physics, in
particular, it can be 
directly related to the $\mathbb C P^1$ nonlinear sigma 
model {\bf [4]}. In fact, solutions of one system can be
transformed into solutions of the other. Moreover
both the generalized Weierstrass system and 
the nonlinear sigma model are completely integrable
systems {\bf [5]}. These types of representations
are useful because in this context, deformations
of surfaces can be discussed in a straightforward way. 
Konopelchenko made
an important remark in this regard. If a
surface is represented locally by the generalized
Weierstrass representation, then based on the
operator
$$
L = \left(  
\begin{array}{cc}
\partial  &  -p  \\
p      & \bar{\partial}  \\
\end{array}   \right)
$$
such that the potential $p$ satisfies a particular evolution equation,
the deformation in $p$ obtained under the evolution equation
induces a local deformation of a surface.
The relevant integrable evolution equation 
considered here is the Novikov-Veselov equation,
although other integrable systems could be considered.
Taimanov {\bf [6]} showed that these formulas for
inducing surfaces in $\mathbb R^3$ describe all
surfaces and that the modified Novikov-Vesolov equation
deforms tori into tori preserving the Willmore functional.
These types of application
lend an important role to this type
of representation.
The analogous problems for surfaces in $\mathbb R^4$ show
that this case is very different from the three-dimensional
case. The main reason is that for tori in $\mathbb R^4$,
each equation of the Davey-Stewartson (DS) hierarchy
describes not one but infinitely many geometrically 
different soliton deformations, and is linked to the fact 
that the representation is not unique in this case.

To put it concisely, a surface in $\mathbb R^3$ is
constructed from a single vector or spinor function ${\bf \psi}$
which is a lift of the Gauss map into nonvanishing
spinors, such that it satisfies a Dirac equation, and
the lift is defined up to a sign.
On the other hand, a surface in $\mathbb R^4$ is
obtained from two spinors ${\bf \psi}$ and
${\bf \varphi}$ which form again a lift of the Gauss
map. However, in this case the lift is defined
only up to a gauge transformation {\bf [7-10]} given
by $e^f$, where $f$ is any smooth function.

It is the purpose in this paper to introduce
the system of equations satisfied by these two
spinors in $\mathbb R^4$, and the corresponding
equations for inducing the corresponding surface.
The DS-hierarchy will be introduced and a
mechanism for deforming these surfaces
will be considered. In particular,
a recipe will be given for obtaining the 
relevant DS equations pertaining to the first
three elements of the hierarchy from a matrix
system using symbolic manipulation {\bf [11]}. 
Applications of these evolution equations will be 
given for deformation of surfaces, 
and some results related to the
Weierstrass representation and its gauge
invariance will be explored. Finally, it will be
shown that the Willmore functional is preserved
under this kind of deformation.

\begin{center}
{\bf 2. REPRESENTATION OF SURFACES AND THEIR DEFORMATIONS}
\end{center}

An extension of the generalized Weierstrass representation
to four-dimensional Euclidean space is based on
a pair of spinors ${\bf \psi}$ and ${\bf \varphi}$ whose
components can be regarded as two independent solutions
of the generalized Weierstrass system in $\mathbb R^3$,
namely (1.1) and (1.2) {\bf [12]}.
Let the spinor functions $\psi = (\psi_1, \psi_2)$ 
and $\varphi= ( \varphi_1, \varphi_2)$ be defined
in a simply connected domain $W \subset \mathbb C$,
parametrized by the complex variable $z$, then the
components each satisfy the pair of Dirac equations
$$
{\cal D} \psi = 0, 
\qquad
\tilde{\cal D} \varphi = 0
\eqno(2.1)
$$
where ${\cal D}$ and $\tilde{\cal D}$ are
matrix operators which are given by
$$
{\cal D} = \left(
\begin{array}{cc}
p    & \partial   \\
- \bar{\partial}  &  \bar{p}  \\
\end{array}   \right),
\qquad
\tilde{\cal D} = \left(
\begin{array}{cc}
\bar{p}  &  \partial   \\
- \bar{\partial}  &  p  \\
\end{array}   \right).
\eqno(2.2)
$$
Before stating how surfaces can be induced from
solutions to system (2.1), the following lemma will
be useful.

{\bf Lemma 1.} (a) The components of the spinors 
$\psi$ and $\varphi$ which satisfy (2.1) also
satisfy the following conditions
$$
\partial ( \varphi_2 \psi_2) + \bar{\partial}
( \varphi_1 \psi_1 ) = 0,
\quad
\bar{\partial} ( \psi_1 \bar{\varphi}_2) 
- \partial ( \bar{\varphi}_1 \psi_2 ) = 0.
\eqno(2.3)
$$

(b) The one-forms defined by 
$$
\eta_k = f_k \, dz + \bar{f}_k \, d \bar{z}
\qquad 
k=1,2,3,4,
\eqno(2.4)
$$
where the coefficients $f_k$ are given by
$$
\begin{array}{ccc}
f_1 = \frac{i}{2} ( \bar{\varphi}_2 \bar{\psi}_2 
+ \varphi_1 \psi_1 ),  &   &  f_2 = \frac{1}{2}
( \bar{\varphi}_2 \bar{\psi}_2 - \varphi_1 \psi_1 ),  \\
     &                   \\
f_3 = \frac{1}{2} ( \bar{\varphi}_2 \psi_1 + \varphi_1 \bar{\psi}_2),
&   &  f_4 = \frac{i}{2} ( \bar{\varphi}_2 \psi_1 - \varphi_1 
\bar{\psi}_2 ).  \\
\end{array}
\eqno(2.5)
$$
are closed.
 
{\em Proof: } (a) Expanding the derivatives in (2.3)
and substituting (2.1) we obtain that
$$
\partial ( \varphi_2 \psi_2 ) + \bar{\partial} ( \varphi_1 \psi_1)
= ( \partial \varphi_2) \psi_2 + \varphi_2 ( \partial \psi_2)
+ (\bar{\partial} \varphi_1) \psi_1 + \varphi_1 ( \bar{\partial} \psi_1 )
$$
$$
=- \bar{p} \varphi_1 \psi_2 - p \varphi_2 \psi_1 + p \varphi_2 \psi_1
+ \bar{p} \varphi_1 \psi_2 =0.
$$
The remaining condition is treated the same way.

(b) Consider the case in which $k=1$, the other
cases proceed in a similar way. Upon using the results from
part (a), we obtain that
$$
d \eta_1 = \bar{\partial} f_1 \, d \bar{z} \wedge dz
+ \partial \bar{f}_1 \, dz \wedge d \bar{z} 
$$
$$
= \frac{i}{2} ( \bar{\partial} ( \bar{\varphi}_2 \bar{\psi}_2)
+ \partial ( \bar{\varphi}_1 \bar{\psi}_1)) \, d \bar{z} \wedge dz
+ \frac{i}{2} ( \partial ( \varphi \psi_2 ) + \bar{\partial}
( \varphi_1 \psi_1 )) d \bar{z} \wedge dz =0.
$$
The next Proposition follows from these.

{\bf Proposition 1.} Let the spinor functions $\psi$ and
$\varphi$ be defined in a simply connected domain $W \subset \mathbb C$ 
and satisfy the Dirac equations (2.1)-(2.2). Then
the one-forms $\eta_k$ in (2.4) define a surface in
$\mathbb R^4$ by means of the integrals
$$
X^k (z, \bar{z} )= X^k (0) + \int_{\Gamma} \eta_k,
\qquad 
k=1,2,3,4.
\eqno(2.6)
$$
The integral in (2.6) is taken over any path $\Gamma$ in $W$.
By Stokes Theorem and Lemma 1, the integral in (2.6)
does not depend on the choice of path.  $\clubsuit$

The induced metric equals
$$
e^{2 \alpha} \, dz \, d \bar{z} =
(| \psi_1|^2 + |\psi_2|^2 )(|\varphi_1|^2 + |\varphi_2|^2) \,
dz \, d \bar{z} = u_1 u_2 \, dz \, d \bar{z},
\eqno(2.7)
$$
where $u_1 = |\psi_1|^2 + |\psi_2|^2$ and $u_2 =
|\varphi_1|^2 + |\varphi_2|^2$. The mean
curvature vector is obtained by calculating
$$
{\bf H} = \frac{2}{e^{2u}} {\bf X}_{z \bar{z}},
\eqno(2.8)
$$
and the norm of the mean curvature vector
is related to $p$ which appears in matrices (2.2)
through the expression
$$
|p| = \frac{e^{\alpha}}{2} | {\bf H}|.
\eqno(2.9)
$$
For $p = \bar{p}$ and $\psi = \pm \varphi$,
these formulas reduce to the generalized Weierstrass
representation for surfaces in $\mathbb R^3$.

In $\mathbb R^3$, a single spinor function
is sufficient to obtain a surface. In this case the
spinor is a lift of the Gauss mapping into nonvanishing
spinors, and is required to satisfy a Dirac equation.
In $\mathbb R^4$, two spinor functions are
required to construct a surface, and these 
functions will form a lift of the Gauss map.
In fact, not every lift actually satisfies the
Dirac equations (2.1). The lifts which do
are defined only up to gauge transformations.

One of the reasons for having this type
of formalism available to generate surfaces
is that deformations of surfaces can be obtained 
and studied in a rigorous way. This constitutes
a very useful application of  these inducing
mechanisms. Deformations of a surface are obtained
by deforming the potential function $p$ which
appears in matrices (2.2) according to some given
evolution equation. In particular, we will be
interested in considering evolution equations which
belong to the DS hierarchy, as these will appear
out of the methodology in due course.
To this end, let us begin to generate these
evolution equations by introducing a corresponding
formalism which produces them in a precise way.

To this end, let the operator $L$ be defined
as follows
$$
L = \left(
\begin{array}{cc} 
0  & \partial   \\
- \bar{\partial}  &  0  \\
\end{array}   \right)
+  \left(
\begin{array}{cc}
-p  &  0   \\
0   &  q   \\
\end{array}  \right).
\eqno(2.10)
$$
We consider deformations of the operator
$L$ which take the form of a triple of
operators $L$, $A$ and $B$. These operators
satisfy
$$
L_t + [ L, A_n ] - B_n L = 0.
\eqno(2.11)
$$
Here $t$ will play the role of the evolution parameter.
Now for any nonzero spinor $\Psi$ we have
$$
[ L , \partial_t - A_n ] \Psi
+ B_n L \Psi =
L ( \partial_t - A_n) \Psi - ( \partial_t - A_n ) L \Psi
+ B_n L \Psi
$$
$$
= L \partial_t \Psi - L A_n \Psi - (\partial_t L) \Psi
- L \partial_t \Psi + A_n L \Psi + B_n L \Psi
$$
$$
= (- \partial_t L - [ L, A_n ] + B_n L ) \Psi.
$$
Therefore, we conclude that (2.11) implies that
$$
[ L, \partial_t - A_n ] + B_n L = 0.
\eqno(2.12)
$$
Now if $L$ satisfies (2.11), then the solution
of the equation
$$
L \Psi = 0
\eqno(2.13)
$$
is evolved according to the equation
$$\Psi_t = A_n \Psi.
\eqno(2.14)
$$

{\bf Theorem 1.} For the case in which $n=1$
such that the matrices $A_1$ and $B_1$ in (2.12)
are given by
$$
A_1 = \left(
\begin{array}{cc}
\partial  &  q   \\
p    & \bar{\partial}  \\
\end{array}   \right),
\qquad
B_1 =  \left(
\begin{array}{cc}
\bar{\partial} -  \partial   &  -p - q  \\
-p  - q  & \partial - \bar{\partial}    \\
\end{array}   \right),
\eqno(2.15)
$$
then (2.12) is exactly equivalent to the
Davey-Stewartson II equations
$$
\partial_t p = \partial p +
\bar{\partial} p,
\qquad
\partial_t q = \partial q + \bar{\partial} q.
\eqno(2.16)
$$

{\bf Proof:} Let $\Psi = ( \psi_1, \psi_2)$ be an arbitrary
two-component spinor, then by matrix operations we have
$$
L \Psi = \left(
\begin{array}{c}
\partial \psi_2 - p \psi_1   \\
- \bar{\partial} \psi_1 + q \psi_2  \\
\end{array}     \right).
\eqno(2.17)
$$
Then based on these matrices,  we calculate
$$
( \partial_t - A_1 ) L \Psi = \partial_t 
\left(
\begin{array}{cc}
- p \psi_1 + \partial \psi_2   \\
- \bar{\partial} \psi_1 + q \psi_2  \\
\end{array}    \right)
 - \left(
\begin{array}{cc}
\partial   &  q  \\
p    &  \bar{\partial}  \\
\end{array}   \right)
\left(
\begin{array}{cc}
-p \psi_1 + \partial \psi_2  \\
- \bar{\partial} \psi_1 + q \psi_2  \\
\end{array}   \right)
$$
$$
\dss =  \left(
\dss \begin{array}{cc}
\partial_t ( -p \psi_1 + \partial \psi_2) - 
\partial ( -p \psi_1 + \partial \psi_2 ) -q 
( - \bar{\partial} \psi_1 + q \psi_2)   \\
\partial_t ( - \bar{\partial} \psi_1 + q \psi_2)
-p ( -p \psi_1 + \partial \psi_2) - \bar{\partial}
(- \bar{\partial} \psi_1 + q \psi_2)    \\
\end{array}   \right)
\eqno(2.18)
$$
as well as
$$
\dss L (\partial_t - A_1 ) \Psi = \left(  \dss
\begin{array}{cc} 
-p  &  \partial  \\
- \bar{\partial}  &  q  \\
\end{array}   \right)
\left(
\begin{array}{cc}
\partial_t \psi_1 - \partial \psi_1 -q \psi_2  \\
\partial_t \psi_2 - p \psi_1 - \bar{\partial} \psi_2  \\
\end{array}   \right)
$$
$$
\dss = \left(
\begin{array}{cc}
-p ( \partial_t \psi_1 - \partial \psi_1 -q \psi_2 ) 
+ \partial ( \partial_t \psi_2 -p \psi_1 - \bar{\partial} \psi_2 )  \\
- \bar{\partial} ( \partial_t \psi_1 - \partial \psi_1 -q \psi_2)
+ q ( \partial_t \psi_2 -p \psi_1 - \bar{\partial} \psi_2)    \\
\end{array}   \right)
\eqno(2.19)
$$
with
$$
B_1 L \Psi = \left(
\begin{array}{cc}
\bar{\partial} - \partial  &  -p-q   \\
-p-q  & \partial - \bar{\partial}    \\
\end{array}   \right)
\left(
\begin{array}{cc}
-p \psi_1 + \partial \psi_2  \\
- \bar{\partial} \psi_1 + q \psi_2  \\
\end{array}   \right)
$$
$$
=  \left(
\begin{array}{cc}
(\bar{\partial} - \partial)(-p \psi_1 + \partial \psi_2)
- (p+q)(- \bar{\partial} \psi_1 + q \psi_2 )   \\
-(p+q)(-p \psi_1 + \partial \psi_2) + ( \partial -\bar{\partial} )
( - \bar{\partial} \psi_1 + q \psi_2 )    \\
\end{array}    \right).
\eqno(2.20)
$$
Substituting (2.18), (2.19) and (2.20) into (2.12)
and simplifying, the top element of the resulting
matrix reduces to
$$
\psi_1 ( \partial_t p - \partial p - \bar{\partial} p ) =0,
$$
and the lower element of the matrix reduces to
$$
\psi_2 ( - \partial_t q + \partial q + \bar{\partial} q ) =0.
$$
These results are exactly system (2.16). $\clubsuit$

These calculations can be best carried out by
means of symbolic manipulation.
In fact, for the cases $n=2$ and $n=3$, the
basic structure of the matrices will be obtained,
and then the rest of the proof makes use of this {\bf [11]}.

It should be noted that the DS I hierarchy is a
related system of nonlinear equations which are
obtained from the DS II hierarchy by replacing
the variables $z$ and $\bar{z}$ by real-valued
variables $x$ and $y$.

Consider the following reduction of the system (2.16)
which is specified by taking
$$
p = -u,
\qquad
q = \bar{u}.
\eqno(2.21)
$$
It is then seen that system (2.16) is compatible under
these substitutions and the pair reduces to the single
expression
$$
\partial_t u = \partial u + \bar{\partial} u.
\eqno(2.22)
$$

{\bf Theorem 2.} For the case in which $n=2$, the matrices
$$
A_2 = \left(
\begin{array}{cc}
- \partial^2 - v_1  &  q \bar{\partial} - \bar{\partial} q  \\
-p \partial + \partial p  &  \bar{\partial}^2 + v_2     \\
\end{array}     \right)
\eqno(2.23)
$$
and 
$$
B_2 = \left(
\begin{array}{cc}
\partial^2 + \bar{\partial}^2 + v_1 + v_2   &  - (p+q) \bar{\partial}
+ \bar{\partial} q - 2 \bar{\partial} p   \\
(p+q) \partial - \partial p + 2 \partial q  &  - (\partial^2
+ \bar{\partial}^2 ) - (v_1 + v_2)         \\
\end{array}     \right)
\eqno(2.24)
$$
where $v_1$ and $v_2$ satisfy
$$
\bar{\partial} v_1 = - 2 \partial (pq),
\qquad
\partial v_2 = - 2 \bar{\partial} (pq)
\eqno(2.25)
$$
generate the following system of equations under (2.12)
$$
\partial_t p = \partial^2 p + \bar{\partial}^2 p 
+ (v_1 + v_2) p,
\qquad
\partial_t q = - \partial^2 q - \bar{\partial}^2 q - 
(v_1 + v_2) q.
\eqno(2.26)
$$

{\bf Proof:} With $L \Psi$ given by (2.17), we have that
$$
( \partial_t - A_2) L \Psi =  \left(
\begin{array}{c}
\partial_t (L \Psi)_1 + \partial^2 (L \Psi)_1 + v_1 (L \Psi)_1
- q \bar{\partial} (L \Psi)_2 + (\bar{\partial} q)(L \Psi)_2    \\
\partial_t (L \Psi)_2 + p \partial (L \Psi)_1 - \partial p
(L \Psi)_1 - \bar{\partial}^2 (L \Psi)_2 - v_2 (L \Psi)_2    \\
\end{array}   \right)
\eqno(2.27)
$$
and 
$$
(\partial_t - A_2) \Psi =  \left(
\begin{array}{c}
\partial_t \psi_1 + \partial^2 \psi_1 + v_1 \psi_1 -q \bar{\partial} \psi_2 
+ \bar{\partial} q \psi_2             \\
\partial_t \psi_2 +p \partial \psi_1 - \partial p \psi_1 -
\bar{\partial}^2 \psi_2 - v_2 \psi_2    \\
\end{array}    \right)
= \left(
\begin{array}{c}
\Lambda_1   \\
\Lambda_2   \\
\end{array}    \right)
\eqno(2.28)
$$
$$
L ( \partial_t - A_2) \Psi =  \left(
\begin{array}{c}
-p \Lambda_1 + \partial \Lambda_2  \\
- \bar{\partial} \Lambda_1 + q \Lambda_2  \\
\end{array}   \right).
\eqno(2.29)
$$
Substituting (2.27), (2.28) and (2.29) into (2.12),
the pair of equations in (2.26) is generated by calculation,
with the $p$ equation the upper component
and the $q$ equation as the lower component. $\clubsuit$

If we consider the reduction given in (2.21),
the $q$ equation in (2.26) is not compatible
with the $p$ equation. However, by modifying the
pair of matrices in a straightforward way, it is
possible to obtain a compatible pair from (2.12).
The following result formalizes this objective.

{\bf Corollary 1.} If the matrices $(A_2, B_2)$
in Theorem 2 are replaced by the matrices 
$( i A_2, i B_2)$, where $A_2$ and $B_2$ are
given by (2.23)
and (2.24) respectively, then (2.12) generates
the following pair of equations
$$
p_t = i ( \partial^2 p + \bar{\partial}^2 p
+ ( v_1 + v_2 ) p ),
\qquad
q_t = - i ( \partial^2 q + \bar{\partial}^2 q
+ (v_1 + v_2 ) q).
\eqno(2.30)
$$
Given Corollary 1, equations (2.30) are
compatible under the reduction given in (2.21),
which reduces the pair of equations given in (2.30)
to the single equation
$$
\partial_t  u = i ( \partial^2 u + \bar{\partial}^2 u 
+ ( v + \bar{v} ) u ),
\qquad
\bar{\partial} v = \partial (|u|^2).
\eqno(2.31)
$$
The last system in the hierarchy to be considered
here is presented in the next result.

{\bf Theorem 3.} For the case in which $n=3$,
define the matrices
$$
A_3 = \left(
\begin{array}{cc}
\partial^3 + \frac{3}{2} v_2 \partial - 3 w_1  &
q \bar{\partial}^2 - (\bar{\partial} q) \bar{\partial}
+ \bar{\partial}^2 q + \frac{3}{2} v_2 q  \\
p \partial^2 - (\partial p) \partial + \partial^2 p
+ \frac{3}{2} v_1 p   &  \bar{\partial}^3 + \frac{3}{2}
v_2 \bar{\partial} - 3 w_2   \\
\end{array}   \right)
\eqno(2.32)
$$
and
$$
B_3 = \left(
\begin{array}{cc}
b_{11}  &  b_{12}   \\
b_{21}  &  b_{22}   \\
\end{array}   \right)
\eqno(2.33)
$$
where
$$
b_{11} = - b_{22} = \bar{\partial}^3 - \partial^3 - \frac{3}{2}
(v_1 \partial - v_2 \bar{\partial}) + 3 ( w_1 - w_2 ),
$$
$$
b_{12} =- (p+q) \bar{\partial}^2 - \frac{3}{2} (p+q) v_2 
- ( 3 \bar{\partial} p - \bar{\partial} q) \bar{\partial}
- ( 3 \bar{\partial}^2 p + \bar{\partial}^2 q ),
\eqno(2.34)
$$
$$
b_{21} = - (p+q)  \partial^2 - \frac{3}{2} (p+q) v_1 -
( 3 \partial q - \partial p )\partial - ( 3 \partial^2 q 
+ \partial^2 p).
$$
such that the $v_i$ and $w_i$ satisfy
$$
\bar{\partial} v_1 =-2 \partial (pq), 
\qquad
\partial v_2 = -2 \bar{\partial} (pq),
\qquad
\bar{\partial} w_1 = \partial (p \partial q),
\qquad
\partial w_2 = \bar{\partial} (q \bar{\partial} p).
\eqno(2.35)
$$
Then (2.12) reduces to the following pair of equations in
terms of $p$ and $q$,
$$
\begin{array}{c}
\partial_t p = \partial^3 p + \bar{\partial}^3 p + \frac{3}{2}
( v_1 \partial p + v_2 \bar{\partial} p)
+ 3 ( w_1 - w_2 + \frac{1}{2} \partial v_1) p,    \\
  \\
\partial_t q = \partial^3 q + \bar{\partial}^3 q + \frac{3}{2}
(v_1 \partial q + v_2 \bar{\partial} q ) 
- 3 ( w_1 - w_2 - \frac{1}{2} \partial v_2 ) q.
\end{array}
\eqno(2.36)
$$
This system can be put in the following form by
redefining $v_i$
$$
\begin{array}{c}
\partial_t p = \partial^3 p + \bar{\partial} p^3
+ 3 ( v_1 \partial p + v_2 \bar{\partial} p )
- 3 ( \partial^{-1} [(q \bar{\partial} p)_{\bar{z}}]
+ \bar{\partial}^{-1} [ (q \partial p)_z]) p,   \\
    \\
\partial_t q = \partial^3 q + \bar{\partial}^3 q +
3 (v_1 \partial q + v_2 \bar{\partial} q )
- 3 ( \partial^{-1} [ (p \bar{\partial} q)_{\bar{z}}]
+ \bar{\partial}^{-1} [ (p \partial q)_z ] )q   \\
\end{array}
\eqno(2.37)
$$
The equations in (2.37) are compatible under
(2.21) and cause (2.37) to reduce to the
single equation
$$
\partial_t u = \partial^3 u + \bar{\partial}^3 u
+ 3 ( v \partial u + \bar{v} \bar{\partial} u)
+ 3 ( w + w') u,
$$
$$
\bar{\partial} v = \partial ( |u|^2 ),
\qquad
\bar{\partial} w = \partial ( \bar{u} \partial u ),
\qquad
\partial w' = \bar{\partial} (\bar{u} \bar{\partial} u ).
\eqno(2.38)
$$
Frequently, (2.31) and (2.38) are referred to as the $DS_{2}$ and
$DS_3$ equations, respectively. In fact, (2.38) is also
compatible with the additional constraint $u = \bar{u}$, and reduces
to the modified Novikov-Veselov equation.
In fact, since $\bar{\partial} v = \partial (u^2)$ and
$\bar{\partial} w = \frac{1}{2} \partial (\partial u^2)= \frac{1}{2} \partial ( \bar{\partial} v) =
\frac{1}{2} \bar{\partial} ( \partial v)$ it follows that
$w = \frac{1}{2} \partial v$ and $w ' = \frac{1}{2} \bar{\partial} v$.
Then (2.38) is given by
$$
\partial_t u = \partial^3 u + \bar{\partial}^3 u 
+ 3 ( v \partial u + \bar{v} \bar{\partial} u )
+ \frac{3}{2} ( \partial v + \bar{\partial} \bar{v})u,
\quad
\bar{\partial} v = \partial (u^2).
\eqno(2.39)
$$
This is exactly the Novikov-Veselov equation.

\begin{center}
{\bf 3. GAUSS MAP, SURFACES AND DEFORMATIONS OF SURFACES IN $\mathbb R^4$.}
\end{center}

The Grassmannian of oriented two-planes in
$\mathbb R^n$ is modeled by the quadric $Q_{n-2}$ in
$C P^{n-1}$ defined by the equation
$$
\sum_{k=1}^n \, z_{k}^2 = 0.
\eqno(3.1)
$$
For any point $P = ( z_1, \cdots, z_n)$ on $Q_{n-2}$
if $z_k = a_k + i b_k$, we obtain a pair of real vectors
$A = ( a_1, \cdots, a_n)$ and $B = ( b_1, \cdots, b_n)$
which satisfy
$$
|A| = |B|,
\quad
A \cdot B = 0.
\eqno(3.2)
$$
These equations are equivalent to (3.1), and $A$ and $B$
cannot be zero since homogeneous coordinates
$( z_1, \cdots, z_n)$ of a point in $CP^{n-1}$ are not
all zero. This pair $A$, $B$ form an orthogonal
basis of an oriented two plane $\Pi$. In $\mathbb R^4$,
we give some facts concerning $Q_2 \subset \mathbb CP^3$.
There are some useful connections between algebraic and
differential geometry which are outlined here {\bf [13,14]}.
The map $\sigma$ from $\mathbb C \times \mathbb C$ given by
$$
\sigma ( w_1, w_2 ) \rightarrow
( 1 + w_1 w_2, i ( 1 - w_1 w_2 ),
w_1 - w_2 , -i ( w_1 + w_2 )),
\eqno(3.3)
$$
has the property that $\sigma^2 = \sum_{k=1}^4 \sigma_k^2 =0$.
Hence $[ \sigma]$ takes values in $Q_2 = \{ [Z] \in \mathbb CP^3 
| Z^2 =0  \}$. Moreover, on $\sigma ( \mathbb C \times \mathbb C)$,
the related mapping $\sigma^{-1}$ is given by
$$
( z_1, z_2, z_3, z_4 ) \rightarrow (w_1, w_2)
= ( \frac{z_3+i z_4}{z_1 -i z_2 },
\frac{-z_3 + i z_4}{z_1 -  i z_2} ).
\eqno(3.4)
$$
Thus, $[\sigma]$ is a biholomorphic map from
$\mathbb C \times \mathbb C$ into $Q_2$, which
when $(w_1, w_2 ) \in \mathbb C \times \mathbb C$
are considered as homogeneous coordinates on
$\mathbb CP^1 \times \mathbb CP^1$. It extends to
a biholomorphic map of $\mathbb CP^1 \times \mathbb CP^1$ onto
$Q_2$. If $\mathbb CP^3$ is considered under the
Fubini-Study metric of constant holomorphic
curvature two, the induced metric on $Q_2$, expressed in
terms of $(w_1 , w_2)$ has the form
$$
ds^2 = \frac{2 |d w_1|^2}{(1 + |w_1|^2)^2} +
\frac{2 |d w_2|^2}{(1 + |w_2|^2)^2}.
\eqno(3.5)
$$
This implies that $Q_2$ is the product of two
standard spheres of constant Gauss curvature of two.

An oriented two-plane in $\mathbb R^4$ is defined
by a positively oriented orthonormal basis
$e_1 = ( e_{1,1}, \cdots, e_{1,4})$ and
$e_2 = ( e_{2,1}, \cdots, e_{2,4})$ defined up
to rotations. There exists a one-to-one
correspondence between components of the $e_i$
and points of the quadric $\mathbb Q \subset \mathbb CP^3$
defined by (3.1) as
$$
z_1^2 + z_2^2 + z_3^2 + z_4^2 =0.
\eqno(3.6)
$$
The correspondence is given by $z_k = e_{1,k}+i e_{2,k}$
for $k=1, \cdots, 4$. Consider
the following change of coordinates from $y_i$ to $z_i$
$$
z_1 = \frac{i}{2} ( y_1 +y_2),
\qquad
z_2 = \frac{1}{2} (y_1 - y_2),
\qquad
z_3 = \frac{1}{2} ( y_3 + y_4),
\qquad
z_4= \frac{i}{2} ( y_3 - y_4).
\eqno(3.7)
$$
The quadric in terms of the $y_i$ can then
be written
$$
y_1 y_2 = y_3 y_4.
$$
This establishes a correspondence between the space
$\mathbb CP^1 \times \mathbb CP^1$ and $\tilde{G}_{4,2}$
in the form of a product
$$
\tilde{G}_{4,2} = \mathbb CP^1 \times \mathbb CP^1.
\eqno(3.8)
$$
It is this equivalence which allows us to decompose the
Gauss map into a pair of maps which can be written as
$$
G = ( G_{\psi}, G_{\varphi} ),   
\quad G : W \rightarrow \tilde{G}_{4,2}   \quad  Q \in W \rightarrow   
( X^1_z (Q), X^2_z (Q) , X^3_z (Q) , X_z^4 (Q)  ),
$$
and $G_{\psi}$ and $G_{\varphi}$ can be represented in terms
of a pair of spinors 
$$
G_{\psi} = ( \psi_1, \psi_2) \in \mathbb CP^1,
\qquad
G_{\varphi} = ( \varphi_1 , \varphi_2 ) \in \mathbb CP^1.
\eqno(3.10)
$$
The actual coordinates of the surface can be
written in terms of the $(\psi_i$ , $\varphi_i)$  
presented in (2.6).

Such a decomposition will however not be unique. In fact, the spinors
$\psi$ and $\varphi$ will be defined only up to a gauge 
transformation {\bf [7-6]}, that is, a transformation of the form,
$$
\left(
\begin{array}{c}
\psi_1   \\
\psi_2   \\
\end{array}  \right) \rightarrow
\left(
\begin{array}{c}
e^f \psi_1   \\
e^{\bar{f}} \psi_2  \\
\end{array}   \right),
\qquad
\left(
\begin{array}{c}
\varphi_1    \\
\varphi_2    \\
\end{array}   \right)  \rightarrow
\left(
\begin{array}{c}
e^{-f} \varphi_1   \\
e^{-\bar{f}} \varphi_2  \\
\end{array}  \right).
\eqno(3.11)
$$
Define the functions $\kappa_{\alpha}$ and
$\tau_{\alpha}$ in terms of $\psi_{\alpha}$
and $\varphi_{\alpha}$, as follows
$$
\begin{array}{cc}
\kappa_1  = e^f \psi_1,  &  \tau_1 = e^{-f} \varphi_1 \\
\kappa_2 = e^{\bar{f}} \psi_2,  &  \tau_2 = e^{-\bar{f}} \varphi_2.  \\
\end{array}
\eqno(3.12)
$$
It is clear that the ratios $\kappa_1 / \bar{\kappa}_2$
and $\tau_1 / \bar{\tau}_2$ are left invariant
by the gauge transformation given by (3.11).

Suppose, for example, a lift is constructed which is
based on the initial pair of functions $(s_1 , s_2 )
= ( e^{i \theta} \cos \eta, \sin \eta )$. Then a
pair of functions $( \psi_1 , \psi_2 )$ can be
sought such that the ratio is preserved, namely,
$\psi_1 /\bar{\psi}_2 = s_1 / s_2$ and 
$(\psi_1, \psi_2)$ satisfy (2.1). From (3.11),
we can write
$$
\psi_1 = e^f s_1,
\qquad
\psi_2 = e^{\bar{f}} s_2.
\eqno(3.13)
$$
Let us obtain equations satisfied by
the functions in the lift as well as $f$.
Requiring that the $\psi_{\alpha}$ in (3.13) satisfy
Dirac equations (2.1), this will be the case provided 
that $( \theta, \eta)$ and $f$ satisfy the
equations
$$
\partial ( e^{\bar{f}} \sin \eta )
+ p e^{f + i \theta} \cos \eta =0,
\qquad
\bar{\partial} ( e^{f + i \theta} \cos \eta )
= \bar{p} e^{\bar{f}} \sin \eta.
\eqno(3.14)
$$
Eliminating the function $p$ from the equations in (3.14), we obtain
$$
\bar{\partial} f + i (\bar{\partial} \theta) \cos^2 \eta =0.
$$
Using (3.14) to obtain $\partial \bar{f}$, 
we obtain an expression for $p$
$$
p = - e^{\bar{f} - f - i \theta} ( i \partial \theta
\sin \eta \, \cos \eta + \partial \eta).
$$
Once the $\psi_{\alpha}$ have been fully determined, the
conditions (2.3) can be used to calculate the components
of the second lift $G_{\varphi}$, since these components
must satisfy the remaining equation in (2.1).

In fact, for a certain class of functions $f$,
the entire system (2.1) may be preserved in form, or may 
be said to be gauge invariant.

{\bf Proposition 2. } If the gauge function $f$
in gauge transformation (3.11) satisfies
$\bar{\partial} f = 0$, then system (2.1)
is left invariant  under (3.11) provided
that the potential function $p$ in  (2.2)
is transformed or gauged according to
$P = p e^{\bar{f}-f}$.

{\bf Proof:} Suppose the functions $\psi_{\alpha}$
and $\varphi_{\alpha}$ satisfy system (2.1).
Differentiating $\kappa_1$ in (3.12) with 
respect to $\bar{z}$, we obtain
$$
\bar{\partial} \kappa_1 =
( \bar{\partial} f) e^f \psi_1 + e^f \bar{\partial} \psi_1
= e^f ( \bar{\partial} f) \psi_1 + \bar{p} e^f \psi_2 
= \bar{P} \kappa_{2}.
$$
A similar result applies to $\kappa_2$. 
Next differentiating $\tau_1$ we have
$$
\bar{\partial} \tau_1 = - ( \bar{\partial} f) e^{-f} 
\varphi_1 + e^{-f} \bar{\partial} \varphi_1
= - e^{-f} ( \bar{\partial} f) \varphi_1 + p
e^{-f} \varphi_2 = P \tau_2.
$$
A similar result holds for the function $\tau_2$. $\clubsuit$

Consider now deformations of surfaces. 
Konopelchenko {\bf [2]} introduced the definition
of the DS deformations of a surface. Integrable
deformations of surfaces generated by the
Weierstrass formulas will be constructed.
As mentioned earlier, this is one of the
main applications of the generalized 
Weierstrass representation, as it gives a
way to construct integrable deformations
of immersed surfaces. 

Let us start with surfaces in $\mathbb R^3$.
Deformations of the functions $\psi$ and $\varphi$
are considered such that there are differential
operators $A_n$, $B_n$, $C_n$ and $D_n$ such that
$$
\psi_{t_n} = A_n \psi + B_n \varphi,
\qquad
\varphi_{t_n} = C_n \psi + D_n \varphi.
\eqno(3.15)
$$
For given operators, the compatibility condition
of (3.15) with (1.1) gives a nonlinear partial differential
equation for $p$. Changing the operators on the right
of (3.15) generates an infinite hierarchy of
integrable equations for $p$.
The deformations of $\psi$ and $\varphi$ described by
(3.15) generates the corresponding deformations
of the corresponding coordinates $X^i (z, \bar{z}, t)$.
For example, when $n=1$, the operators in (3.15)
can be written as
$$
A_1 = \alpha \bar{\partial},
\quad
B_1 = \gamma p,
\quad
C_1 = \alpha q,
\quad
D_1 = \gamma \partial,
\eqno(3.16)
$$
and system (3.15) turns out to be linear.
Equations corresponding to higher values of $n$ 
are nonlinear equations, such as the ones presented 
in Theorems $2$ and $3$.

Now to generate integrable deformations of
surfaces immersed in $\mathbb R^4$,
it is assumed that the components of both spinor
solutions to (2.1) and (2.2) evolve with parameter $t$
according to (3.15) under the same operators
$A_n$, $B_n$, $C_n$ and $D_n$. The compatibility
conditions for (2.1) with (3.15) fix the dependence
of $\psi$ and $\varphi$ as well as $p$ and $q$ on 
the parameter $t$, and consequently define the
deformations of the corresponding surfaces.
Thus, the coordinates $X^i$ for $i=1,2,3,4$
of the given surface which are calculated by means 
of (2.6) are defined in terms of $t$ as well.
Specific cases will be governed by different
reductions of the hierarchy. This can now be
put together and stated in a precise way.
First, let us note that $A_n$ in Theorems 1 to 3
depend on two functional parameters $p$ and $q$ 
and for the case in which $p=-u$ and $q= \bar{u}$, 
define $A_n^{+} = A_n$. For the alternate case,
$p=- \bar{u}$ and $q=u$, set $A_n^{-} = A_n$.
In terms of these new operators, the following Proposition 
can be stated.

{\bf Proposition 3:} Let surface $\Sigma$ be defined
by (2.5) and (2.6) for certain initial spinors
$\psi^0$ and $\varphi^0$ which satisfy (2.1),
and let $p (z, \bar{z}, t)$ be a deformation
of the potential whose evolution is described 
by the reduced equations (2.16), (2.31) or (2.38).
Then the equations (2.5) and (2.6) combined with
a pair of the following equations
$$
\begin{array}{ccc}
\psi_t = A_1^{+} \psi,  &  &  \varphi_t = A_1^{-} \varphi,  \\
    &  &                 \\
\psi_t = i A_2^{+} \psi, &  &  \varphi_t = - i A_2^{-} \varphi,  \\
    &  &                 \\
\psi_t = A_3^{+} \psi,  &  &  \varphi_t = A_3^{-} \varphi,
\end{array}
\eqno(3.17)
$$
such that $\psi_{t=0} = \psi^0$ and $\varphi_{t=0}
= \varphi^0$ define deformations of the surface which
is governed by the reduced evolution equations
(2.16), (2.31) or (2.38), respectively.

{\bf Proof:} The deformation of $p$ in (2.2) is
described by (2.11)-(2.12), so the spinors
$\psi^0$ and $\varphi^0$, which satisfy 
system (2.1), are
deformed according to (3.17). Thus, for any
$t$, the resulting spinors still satisfy the
Dirac equations (2.1). From Proposition 1,
however, solutions to this system define a new 
related surface $\Sigma_t$ in $\mathbb R^4$ 
by means of Weierstrass equations (2.4)-(2.6).
This process has generated a deformed surface
$\Sigma_t$ such that $\Sigma_0$ coincides
with the original surface generated by
spinors $(\psi^0 , \varphi^0 )$.  $\clubsuit$

Finally several results related to preservation of
surface structure under deformation will be given.

{\bf Proposition 4.} Let $h$ be any of the functions
$\bar{\psi}_1 \bar{\varphi}_1$, $\bar{\psi}_1 \varphi_2$,
$\psi_2 \bar{\varphi}_1$, $\psi_2 \varphi_2$ which
appear in the conservation laws (2.3). Then with
respect to $DS_2$ equation (2.31), each of the
functionals
$$
J (h) = \int_{\Sigma} h \, dz \wedge d \bar{z}
\eqno(3.18)
$$
is conserved with respect to the evolution parameter,
$\partial_t J =0$.

{\bf Proof:} The claim will be shown for the
case in which $h = \bar{\psi}_1 \bar{\varphi}_1$,
the others follow similarly. To do this, the
evolution equations for $\bar{\psi}_1$ and
$\bar{\varphi}_1$ are required. These are given
by the second pair of matrix equations in (3.17) 
such that $A_2^{\pm}$ are based on the $A_2$
given in (2.23). These equations are given by
$$
\bar{\psi}_{1t} = -i [ (- \bar{\partial}^2 - \bar{v})
\bar{\psi}_1 + (u \partial - \partial u) \bar{\psi}_2 ],
\qquad
\bar{\varphi}_{1t} = i [ - (\bar{\partial}^2 + \bar{v})
\bar{\varphi}_1 + (\bar{u} \partial - \partial \bar{u} )
\bar{\varphi}_2 ].
\eqno(3.19)
$$
Let
$$
J = \int_{\Sigma} \bar{\psi}_1 \bar{\varphi}_1 \, dz
\wedge d \bar{z},
$$
then differentiation proceeds through the integral to give
$$
\partial_t J = \int_{\Sigma} ( \bar{\psi}_{1t}
\bar{\varphi}_1 + \bar{\psi}_1 \bar{\varphi}_{1t} ) \, 
dz \wedge d \bar{z}
$$
$$
=-i \int_{\Sigma} [ ( - \bar{\partial}^2 \bar{\psi}_1
- \bar{v} \bar{\psi}_1 ) \bar{\varphi}_1 + ( u \partial 
\bar{\psi}_2 - (\partial u) \bar{\psi}_2 ) \bar{\varphi}_1
+ \bar{\psi}_1 ( \bar{\partial}^2 \bar{\varphi}_1 + \bar{v}
\bar{\varphi}_1 ) - \bar{\psi}_1 ( \bar{u} \partial
\bar{\varphi}_2 - ( \partial \bar{u}) \bar{\varphi}_2) ]
\, dz \wedge d \bar{z}
$$
$$
=- i \int_{\Sigma} [ ( u \partial \bar{\psi}_2 -
( \partial u) \bar{\psi}_2 ) \bar{\varphi}_1 
- ( \bar{u} \partial \bar{\varphi}_2 - (\partial \bar{u})
\bar{\varphi}_2 ) \bar{\psi}_1 ] \, dz \wedge d \bar{z},
$$
where (3.19) and integration by parts has been
used to simplify this. Integrating by parts once more,
we obtain
$$
\partial_t J = -  i \int_{\Sigma}
[ u \bar{\varphi}_1  \partial \bar{\psi}_2 + u \partial
( \bar{\varphi}_1 \bar{\psi}_2 ) - \bar{u}
\partial (\bar{\varphi}_2 \bar{\psi}_1 ) - \bar{u}
\bar{\psi}_1 \partial  \bar{\varphi}_2  ] \, dz \wedge d \bar{z}
$$
$$
=-i \int_{\Sigma} [ 2u \bar{\varphi}_1 \partial \bar{\psi}_2 
+ u \bar{\psi}_2 \partial \bar{\varphi}_1 - 2 \bar{u} \bar{\psi}_1 
\partial \bar{\varphi}_2  - \bar{u}
\bar{\varphi}_2 \partial \bar{\psi}_1 ] \, dz \wedge d \bar{z}.
$$
Finally, replace both $u$ and $\bar{u}$ in this by the
corresponding derivatives from (2.1) to get
$$
\partial_t J = - i \int_{\Sigma}
[ 2 \bar{\partial} \bar{\varphi}_2 \partial \bar{\psi}_2
+ \partial \bar{\psi}_1 \partial \bar{\varphi}_1
- 2 \bar{\partial} \psi_2 \partial \bar{\varphi}_2
- \partial \bar{\varphi}_1 \partial \bar{\psi}_1 ] \, 
dz \wedge d \bar{z} 
$$
$$
= - 2 i \int_{\Sigma} [ \bar{\partial} \bar{\varphi}_2
\partial \bar{\psi}_2 - \bar{\partial} \bar{\psi}_2
\partial \bar{\varphi}_2 ] \, dz \wedge d \bar{z}
$$
$$
= - 2 i \int_{\Sigma}
[ - \bar{\varphi}_2 \bar{\partial} \partial \bar{\psi}_2
+\bar{\varphi}_2 \partial \bar{\partial} \bar{\psi}_2  ]
\, dz \wedge d \bar{z} =0.
$$
{\bf Proposition 5.} Let $W$ be the Willmore functional
defined as
$$
W = \int_{\Sigma} | u|^2 \, dz \wedge d \bar{z}.
\eqno(3.20)
$$
Then the $DS_2$ deformation (2.31) of tori preserves
the Willmore functional defined by (3.20).

{\bf Proof:} Differentiating $W$ given in (3.20) with respect to
$t$ through the integral , we obtain
$$
\partial_t W = \int_{\Sigma} ( u_t \bar{u} + 
u \bar{u}_t ) \, dz \wedge d \bar{z}.
$$
Replacing the derivatives with respect to $t$
by the $DS_2$ equation (2.31), this becomes
$$
\partial_t W = i \int_{\Sigma} (( \partial^2 u
+ \bar{\partial}^2 u + 2 ( v  + \bar{v}) u ) \bar{u}
- u ( \bar{\partial}^2 \bar{u} + \partial^2 u 
+ 2 ( v + \bar{v}) \bar{u})) \, dz \wedge d \bar{z}
$$
$$
= i \int_{\Sigma} ( \bar{u} \partial^2 u - u \partial^2 \bar{u}
+ \bar{u} \bar{\partial}^2 u - u \bar{\partial}^2 \bar{u}) 
\, dz \wedge d \bar{z}.
$$
Finally, integrating this by parts twice, the required
result is obtained
$$
\partial_t W = i \int_{\Sigma}
( \bar{u} \partial^2 u - ( \partial^2 u) \bar{u}
+ \bar{u} \bar{\partial}^2 u
- ( \bar{\partial}^2 u) \bar{u} ) \,
dz \wedge d \bar{z} =0.
$$

\newpage
\begin{center}
{\bf REFERENCES.}
\end{center}

\noindent
$[1]$ Konopelchenko B. G., Introduction to 
Multidimensional Integrable Equations, Plenum
Press, New York, 1992.   \\
$[2]$ Konopelchenko B. G., Induced Surfaces and
their Integrable Dynamics, Studies in Appl. Math.,
1996, {\bf 96}, 9-51.    \\
$[3]$ Konopelchenko B. G. and Taimanov I. A.,
Constant Mean Curvature Surfaces via an
Integrable Dynamical System, J. Phys. A:
Math. Gen., 1996, {\bf 11,7}, 1183-1216.  \\
$[4]$ Bracken P., Grundland A. M. and Martina L.,
The Weierstrass-Enneper
System for Constant Mean Curvature Surfaces
and the Completely Integrable Sigma Model,
J. Math. Phys., 1999, {\bf 40}, 3379-1403.  \\
$[5]$ Bracken P. and Grundland A. M., On Complete
Integrability of the Generalized Weierstrass
System, J. of Nonlinear. Math. Phys., 2002, {\bf 9,2}, 229-247.   \\
$[6]$ Taimanov I. A., The Weierstrass representation
of closed surfaces in $\mathbb R^3$, Funct. Anal. Appl.,
1998, {\bf 32}, 258-267.   \\
$[7]$ Bracken P and Grundland A. M., Solutions of the
Generalized Weierstrass Representation in Four-Dimensional
Euclidean Space, J. of Nonlinear Math. Phys., 2002, {\bf 9,3},
357-381.  \\
$[8]$ Taimanov I. A., Surfaces in the four-space and
the Davey-Stewartson equations, J. of Geometry and Physics,
2006, {\bf 56}, 1235-1256.   \\
$[9]$ Taimanov I. A., Modified Novikov-Veselov equation
and differential geometry of surfaces, Am. Math.
Soc. Transl. Ser. 2 1997, {\bf 179}, 133-151.   \\
$[10]$ Taimanov I. A., Surafces of revolution in
terms of solitons, Ann. Global Anal. Geom.,
1997, {\bf 12}, 419-435.  \\
$[11]$ Char B. W., Geddes K. O., Leong B. L.,
Monagan M., Watt S., Maple V, Language Reference Manual,
Springer, New York, 1991.   \\
$[12]$ Konopelchenko B. L. and Landolfi G.,
Generalized Weierstrass Representation for
Surfaces in Multidimensional Riemann Spaces,
J. of Geom. Physics, 1999, {\bf 29}, 319-333.  \\
$[13]$ Hoffman D. A. and Osserman R.,
The Gauss Map of Surfaces in $\mathbb R^n$,
J. Differential Geometry, 1983, {\bf 18}, 733-754.  \\
$[14]$ Hoffman D. A. and Osserman R., The Gauss 
Map of Surfaces in $\mathbb R^3$ and $\mathbb R^4$,
Proc. London Math. Soc., 1985, {\bf 50}, 27-56.  \\
\end{document}